\documentclass[conference]{IEEEtran}
\IEEEoverridecommandlockouts
\usepackage{cite}
\usepackage{amsmath,amssymb,amsfonts}
\usepackage{algorithmic}
\usepackage{graphicx}
\usepackage{textcomp}
\usepackage{xcolor}
\usepackage[a4paper, total={184mm,239mm}]{geometry}

\usepackage{hyperref}
\usepackage{booktabs}
\usepackage{multirow}
\usepackage{xspace}
\newcommand{\archname}{MILLION\xspace}

\DeclareMathOperator*{\concat}{\scalerel*{\Vert}{\sum}}
\usepackage[capitalise]{cleveref}

\usepackage{scalerel}
\usepackage{tikz}
\newlength\szg     \newcommand \hquan[1]{%
\settoheight\szg{#1}%
\tikz[baseline]{
\pgfmathparse{1}
\let\hfs\pgfmathresult
\filldraw (0,\szg/2) circle (\szg/2+0.35ex);
\node[white] at (0,\szg/2) {\makebox[0em][c]{\scalebox{\hfs}[1]{\textbf{#1}}}};
}}

\newcommand \hbquan[1]{%
\settoheight\szg{#1}%
\tikz[baseline]{
\pgfmathparse{1}
\let\hfs\pgfmathresult
\filldraw[fill=white, draw=black] (0,\szg/2) circle (\szg/2+0.35ex);
\node[black] at (0,\szg/2) {\makebox[0em][c]{\scalebox{\hfs}[1]{\textbf{#1}}}};
}}
\usepackage{booktabs}
\usepackage{multirow}

\def\BibTeX{{\rm B\kern-.05em{\sc i\kern-.025em b}\kern-.08em
    T\kern-.1667em\lower.7ex\hbox{E}\kern-.125emX}}
\begin{document}
\title{\textbf{MILLION:} \underline{M}aster\underline{I}ng \underline{L}ong-Context \underline{L}LM \underline{I}nference Via \underline{O}utlier-Immunized KV Product Qua\underline{N}tization \\

}
\author{
\IEEEauthorblockN{Zongwu Wang$^{1,2,\dagger}$, Peng Xu$^{1,2,\dagger}$, Fangxin Liu$^{1,2,*}$, Yiwei Hu$^{1}$, Qingxiao Sun$^{4}$, \\ Gezi Li$^{3}$, Cheng Li$^{3}$, Xuan Wang$^{3}$, Li Jiang$^{1,2,*}$, and Haibing Guan$^{1}$}
\IEEEauthorblockA{\textit{1. Shanghai Jiao Tong University, \,\,\,\,\,\,\, 2. Shanghai Qi Zhi Institute,} }
\IEEEauthorblockA{\textit{3. Huawei Technologies Co., Ltd, \,\,\,\,\,\,\, 4.China University of Petroleum-Beijing}\\
\IEEEauthorblockA   {*Corresponding Author \,\,\,\,\,\,\,\{wangzongwu,\,liufangxin,\,ljiang\_cs\}@sjtu.edu.cn}
}
\thanks{$\dagger$ These authors contributed equally. This work was supported by the National Key Research and Development Program of China (2024YFE0204300), National Natural Science Foundation of China (Grant No.62402311), and Natural Science Foundation of Shanghai (Grant No.24ZR1433700).}
}


\maketitle

\begin{abstract}

Large language models (LLMs) are increasingly utilized for complex tasks requiring longer context lengths, with some models supporting up to 128K or 1M tokens. This trend, however, presents significant challenges in inference speed and memory management.
The primary bottleneck in long-context LLM inference is the quadratic computational complexity of attention mechanisms, causing substantial slowdowns as sequence length increases. KV cache mechanism alleviates this issue by storing pre-computed data, but introduces memory requirements that scale linearly with context length, hindering efficient LLM deployment. Quantization emerges as a promising approach to address the widening gap between LLM size and memory capacity. However, traditional quantization schemes often yield suboptimal compression results for KV caches due to two key factors:
i) On-the-fly quantization and de-quantization, causing significant performance overhead;
ii) Prevalence of outliers in KV values, challenging low-bitwidth uniform quantization.
To this end, we propose \archname, a novel quantization framework achieving low-bitwidth KV cache through product quantization. First, we conduct a thorough analysis of KV cache distribution, revealing the limitations of existing quantization schemes. Second, we introduce a non-uniform quantization algorithm based on product quantization, which efficiently compresses data while preserving accuracy. Third, we develop a high-performance GPU inference framework with efficient attention kernel and pipeline design for \archname that leverages sparse computation and asynchronous quantization, significantly enhancing inference speed. 
Comprehensive evaluation results demonstrate that \archname can achieve 4 bits quantization with trivial perplexity and accuracy loss, and achieve 2.09x end-to-end performance gains at 32K context length. Code is released at \url{https://github.com/ZongwuWang/MILLION}.
\end{abstract}


\vspace{-0.2cm}
\section{Introduction}
Large language models (LLMs) have demonstrated remarkable capabilities across various tasks, with recent efforts focusing on expanding their context lengths to further enhance performance.
Longer context lengths enable new applications such as long document summarization, retrieval-based question answering for extensive documents, extended multi-turn interactions, and code analysiscite\cite{chen2023longlora}. This demand has driven significant advancements in long-context models in both industry\cite{Introduc3,ModelsOp33} and academiacite\cite{chen2023longlora}. While increased contextual capability enhances model intelligence, it also introduces substantial computational demands, scaling quadratically ($O(n^2)$) with context length. To address this challenge, modern LLM frameworks have adopted the key-value (KV) cache, which reduces the computational complexity to $O(n)$ at the cost of increased memory usage. However, the practice of batch processing requests, while improving tensor processing efficiency, further exacerbates storage overhead as KV caches cannot be shared between multiple requests\cite{yu2022orca,agrawal2024taming}.

KV cache storage scales linearly with context length, often surpassing device memory limits. For example, in the 540B PaLM model, a batch size of 512 and context length of 2048 can result in a 3TB KV cache—three times the model's parameter size. Additionally, each token generation in the decoding phase requires full KV cache access, creating a major memory bottleneck in long-context scenarios.
Recent research has focused on addressing these memory-related challenges in LLMs through three primary approaches:
\textbf{1) Optimized attention mechanisms:} Techniques such as MQA\cite{shazeer2019fast}, GQA\cite{ainslie2023gqa}, MLA\cite{liu2024deepseek}, YOCO\cite{sun2024you}, and CHAI\cite{agarwal2024chai} have been proposed to reduce computational and storage overheads by sharing KVs among attention heads. However, these optimizations often come at the cost of reduced accuracy, with only GQA currently seeing widespread adoption in state-of-the-art LLMs.
\textbf{2) Sparse attention mechanisms:} These approaches aim to reduce the complexity of attention computations. For example, given the importance of local context, sliding window attention maintains a fixed-size sliding window of recent token states to limit the number of tokens involved in computations\cite{beltagy2020longformer}. Streaming-LLM further improves on this by introducing the ``attention sink'' tokens \cite{xiao2023efficient}, demonstrating that retaining initial token KVs can recover much of the lost performance. More advanced KV sparsity techniques \cite{zhang2024h2o, yang2024pyramidinfer, dai2024corm, li2024snapkv, adnan2024keyformer, ham2021elsa} attempt to identify and remove less important token KVs based on historical attention patterns. However, recent studies \cite{yang2024no, tang2024quest} suggest that these approaches may suffer from accuracy degradation, as past attention distributions may not reliably predict future attention needs in long-context generation.
\textbf{3) Key-value quantization: }Quantization, successful in model weight compression \cite{egiazarian2024extreme,ashkboos2024quarot,frantar2022gptq,lin2024awq}, has emerged as a promising technique for KV cache optimization. However, applying weight quantization methods directly to KV caches presents unique challenges:
First, the presence of outliers in the KV states of LLMs expands the quantization range, making low-bit quantization difficult. Second, since KVs are dynamically generated during inference, quantization or de-quantization delays can negatively impact real-time performance. According to \cite{lin2024qserve}, 4 bit integer dequantization contributes 20\% to 90\% of the inference latency.
Additionally, linear attention mechanisms like Manba\cite{gu2023mamba}, RWKV\cite{peng2023rwkv} and InfiniTransformer\cite{munkhdalai2024leave} have been developed to further reduce the complexity of attention computations. However, these approaches also tend to trade off some degree of model accuracy for efficiency gains.



Recent works have attempted to address these challenges: QServer \cite{lin2024qserve} proposed SmoothAttention, which implements 4-bit KV cache quantization with the SmoothAttention scale $\Lambda$ into previous layers’ weights to reduce de-quantization overhead, however the accuracy is limited by the hard constraint introduced by the rotary positional embedding.
KIVI employs group-wise quantization, adapting to outlier distribution through channel-wise and token-wise quantization of keys and values.
KVQuant further refines this approach with outlier isolation and non-uniform quantization, improving low-bitwidth accuracy at the cost of inference performance. Unfortunately, these techniques often degrade inference speed due to the overhead introduced by outlier isolation and non-uniform quantization. The ideal solution would balance these factors without significantly compromising any single aspect.

This work introduces \archname, a novel quantization scheme for efficient dynamic KV compression in LLMs. \archname utilizes non-uniform quantization and eliminates de-quantization overhead, integrating seamlessly with a high-performance system optimized for GPU-based inference.
First, we thoroughly analyze the distribution patterns of KVs in LLMs, uncovering the limitations of existing quantization methods. Based on these distribution characteristics, we propose a non-uniform quantization scheme using Product Quantization (PQ). By clustering vectors, PQ distributes quantization power unevenly across channels, which helps handle outliers in the KV states. PQ also naturally supports mixed-precision quantization, offering flexibility in balancing performance and accuracy.
Finally, we design an efficient GPU inference framework for \archname, which accelerates inference by leveraging sparse computation and asynchronous quantization, boosting processing speed while reducing memory overhead.

In summary, we make the following contributions:

\begin{itemize}

    \item \textbf{KV distribution analysis}: We perform an in-depth analysis of the distribution characteristics of key-value pairs in LLMs, identifying the limitations of existing quantization methods and highlighting the unique challenges posed by outliers and dynamic KV generation.
    \item \textbf{PQ-based non-uniform quantization}: Building on the insights from the KV distribution analysis, we propose a novel non-uniform quantization scheme based on PQ. This approach leverages clustering to distribute quantization power unevenly across channels, enabling efficient compression while maintaining high accuracy. 
    \item \textbf{Efficient GPU inference system}: We design and implement a high-performance GPU-based inference framework optimized for \archname. This system accelerates inference through sparse computation and asynchronous quantization, effectively eliminating de-quantization overhead while ensuring fast, scalable, and memory-efficient processing of long-context inputs.
\end{itemize}

\section{PRELIMINARY}

\subsection{Attention-based LLM model.}

\begin{figure}[t]
\setlength{\abovecaptionskip}{1pt}  
\setlength{\belowcaptionskip}{1pt} 
\centering
\includegraphics[width=0.95\linewidth]{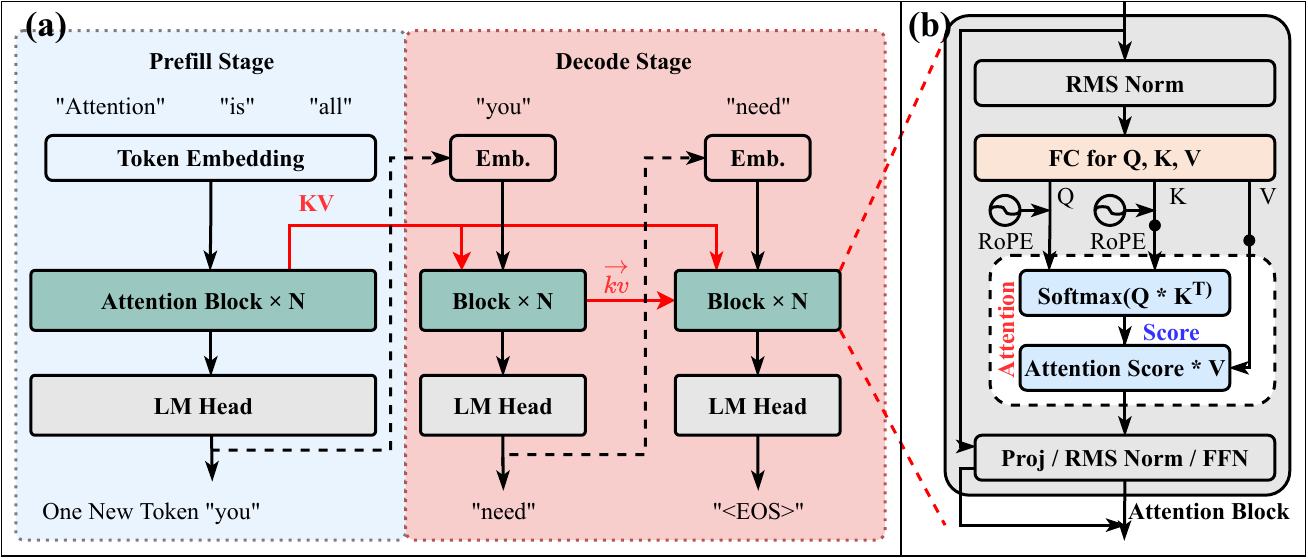}
\caption{Overview of transformer model structure. (a) Transformer model consists of two stage, the prefill stage processes the prompt prefilling of all tokens in batches, and the decode stage generates tokens one by one; (2) Details of the attention block.}
\label{fig:llm}
\vspace{-0.3cm}
\end{figure}
The Transformer model architecture consists of two main stages: the Prefill stage and the Decode stage, as depicted in Fig.~\ref{fig:llm}(a). In the Prefill stage, the model processes the entire input sequence in batches, and generate the key and value vectors for each prompt tokens simultaneously.
The Decode stage operates auto-regressively, generating the output sequence token by token. 
The auto-regressively interaction is handled by the attention blocks, which is the most important modules in the transformer models. 
The detailed algorithm of the attention block is depicted in Fig.~\ref{fig:llm}(b). 
The unique attention mechanism, highlighted with a white background, enables token interactions within the context. This process is formalized as follows:
\begin{equation}
\label{eq:attn}
    A_n = \mathrm{Softmax}(\frac{Q \times K^T}{\sqrt{d_k}}) \times V
\end{equation}

Where $Q \in \mathbb{R}^{n_q \times d_k}$, $K$ and $V \in \mathbb{R}^{n_k \times d_k}$ represent the query, key, and value vectors corresponding to each token, $d_k$ denotes the head dimension, $n_q$ and $n_k$ denote the token number of query and key/value for the attention computation, respectively. Due to the causality of the language model, the query vector of each token needs to be feature fusion with all previous key/value vectors. 
Then the attention scores are utilized to calculate the weighted sum of the value vectors for each token, resulting in the current context state vector. This is followed by feature extraction using additional FFN layers.

\subsection{Integer Quantization}

Existing integer quantization maps high-precision numbers to discrete levels, the process can be formulated as:
\begin{equation}
\label{eq:quant}
\left\{
\begin{array}{l}
    \boldsymbol{Q_X} = \lceil \frac{\boldsymbol{X}}{s} + z \rfloor \\
    s = \frac{\boldsymbol{X}_{max} - \boldsymbol{X}_{min}}{q_{max} - q_{min}} \\
    z = \lceil q_{min} - \frac{\boldsymbol{X}_{min}}{s} \rfloor
\end{array}
\right.
\end{equation}
where $\boldsymbol{X}$ is the floating point tensor, $\boldsymbol{Q_X}$ is its n-bit quantized counterpart, $s$ is the scaling factor and $z$ is the zero point. Thus, the dequantized tensor can be represented as,
\begin{equation}
\label{eq:dequant}
    \hat{\boldsymbol{X}} = Q(\boldsymbol{X}) = (\boldsymbol{Q_X} - z) \cdot s
\end{equation}
This is known as asymmetric quantization, where $\boldsymbol{X}{max} = max(\boldsymbol{X})$, $\boldsymbol{X}{min} = min(\boldsymbol{X})$, and $q_{max} - q_{min} = 2^n - 1$ for n-bit integer quantization. Fig.~\ref{eq:quant} can be further simplified to symmetric quantization, where the zero point $z$ is set to 0, and the range is centered around zero with $\boldsymbol{X}{max} = -\boldsymbol{X}{min} = max |\boldsymbol{X}|$. In this case, $q_{max} - q_{min} = 2^n - 2$, as one bit is reserved for the sign.

The de-quantization process of integer quantization needs to be implemented using general-purpose CUDA cores, which are less efficient for this type of operation. This implementation introduces a significant performance bottleneck, resulting in 20\% to 90\% overhead \cite{lin2024qserve}.

\begin{figure}[t]
\vspace{-1.3cm}
\setlength{\abovecaptionskip}{1pt}  
\setlength{\belowcaptionskip}{1pt} 
\centering
\includegraphics[width=0.95\linewidth]{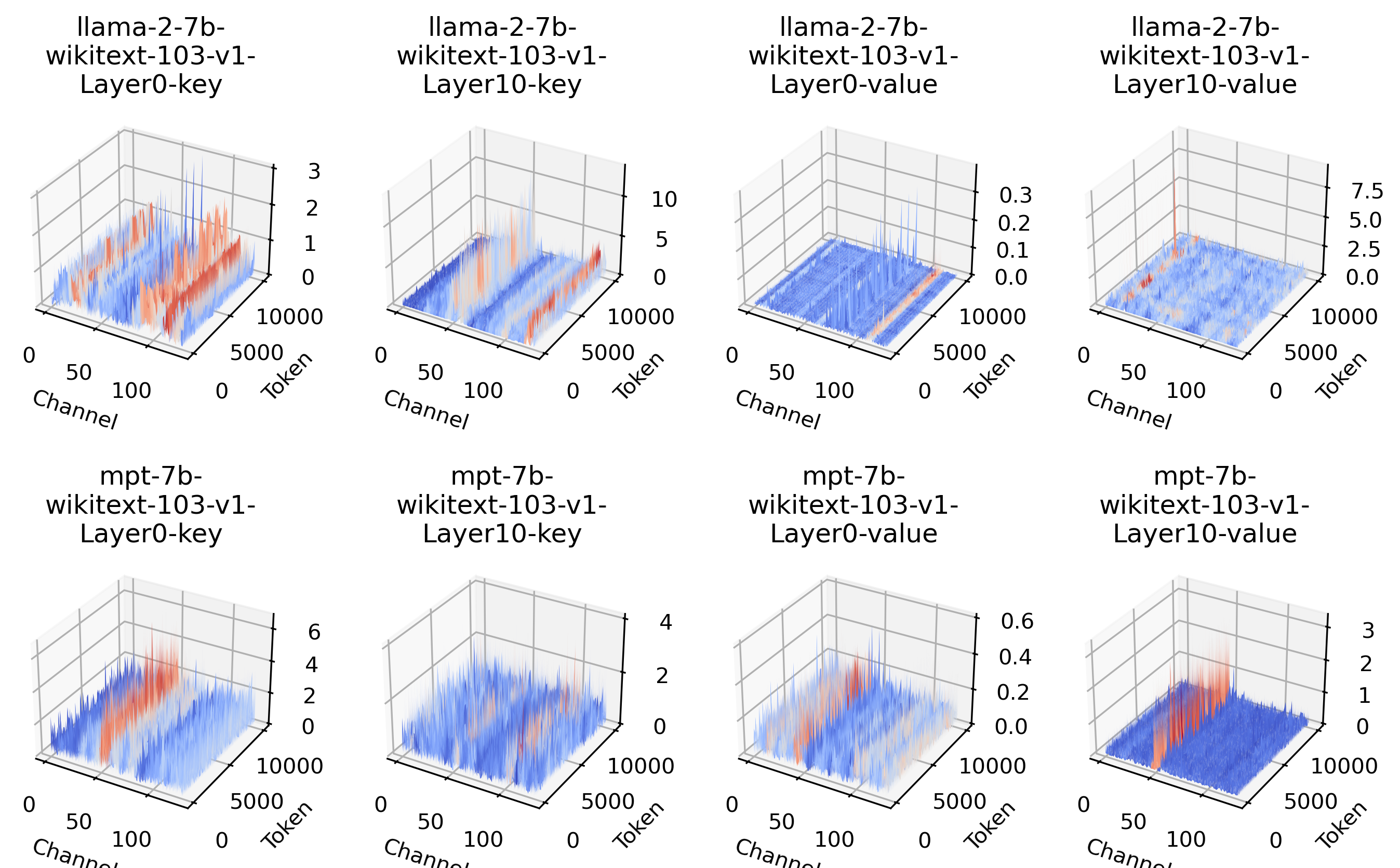}
\caption{Magnitude distribution of key and value cache for Llama-2-13B and Falcon-7B.}
\label{fig:kvdis}
\vspace{-0.3cm}
\end{figure}
\subsection{Product Quantization}
Product Quantization (PQ) is an innovative algorithm that revolutionizes high-dimensional vector retrieval. It offers an efficient and scalable solution for large-scale approximate nearest neighbor searches (ANNS) in high-dimensional spaces\cite{jegou2010product}. The key concept of PQ is the partitioning of high-dimensional vectors into several lower-dimensional subspaces, each independently quantized. This approach significantly reduces the complexity and storage cost of high-dimensional vector similarity operations. According to \cite{simhadri2022results}, PQ can achieve 90\% memory compression without significant accuracy loss.

The PQ process begins with a training phase, where a codebook is created for each subspace using k-means clustering.  Each vector is partitioned and assigned to the nearest centroid in its subspace, with the centroid index stored as a compact code, effectively compressing the data and reducing storage. During the retrieval phase, query vectors are similarly partitioned and quantized. 
The algorithm approximates the distance by summing precomputed distances between the query subvectors and the centroids of each corresponding subspace in the database.
This asymmetric distance computation enables fast ANNS without decoding, achieving both storage efficiency and quick retrieval.

\subsection{Motivation}

In this section, we analyze the distribution of KV cache to reveal the limitations of existing low-precision quantization methods. This insight informs our proposed product quantization-based scheme for efficient KV cache compression.

\textit{1) Outliers hinder low-precision quantization.} To explore the root cause of poor performance in existing KV quantization algorithms, we analyzed the KV distributions across different models, as shown in Fig.~\ref{fig:kvdis}. Our findings reveal that in various models, outliers in the key cache are concentrated in several channels, while the value cache vectors also contain outlier points but without obvious anisotropy in their distribution. As shown in Fig.~\ref{eq:quant}, quantization accuracy depends on the distribution range of the tensor for a given bit width. Concentrated tensor distributions yield higher quantization accuracy. However, outliers significantly expand the numerical distribution range, complicating low-precision quantization. While group-wise and mixed-precision quantization can mitigate outliers, they introduce additional overheads for quantization/de-quantization.

\begin{figure}[t]
\vspace{-1.3cm}
\setlength{\abovecaptionskip}{1pt}  
\setlength{\belowcaptionskip}{1pt} 
\centering
\includegraphics[width=0.95\linewidth]{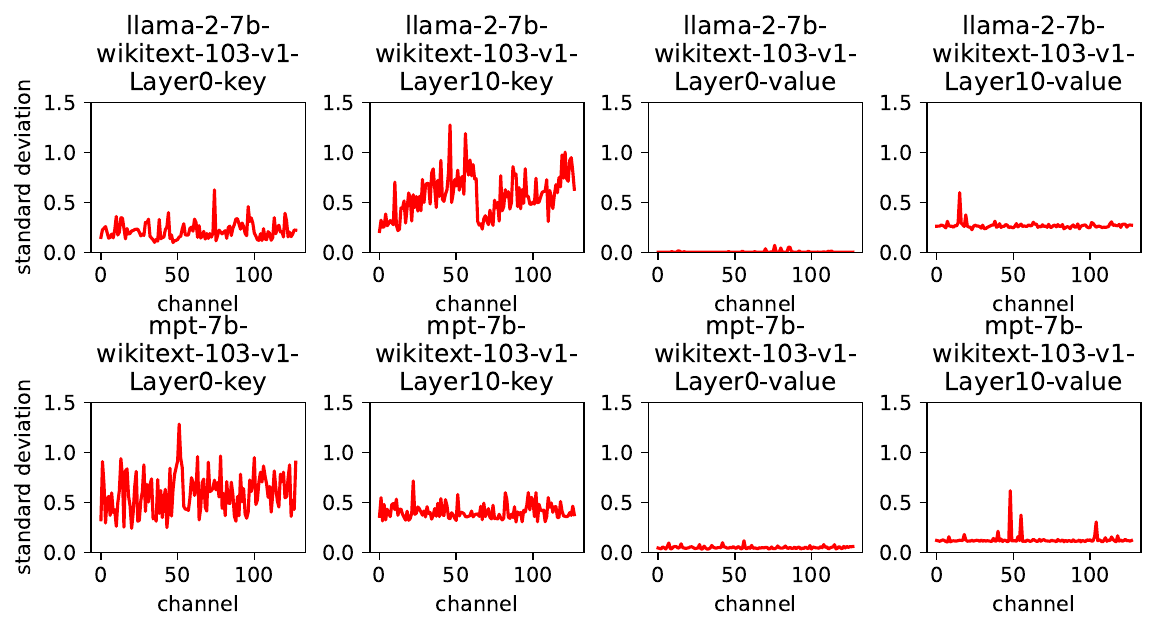}
\vspace{-0.2cm}
\caption{channel-wise standard deviation distribution of key and value cache for Llama-2-13B and Falcon-7B.}
\label{fig:stdis}
\vspace{-0.3cm}
\end{figure}

\textit{2) Exploring Opportunities for Non-uniform Quantization.} We further explore the standard deviation distribution of KV across different channels in \cref{fig:stdis}. Key standard deviations are notably prominent in certain channels, which we term as standard deviation outliers. In contrast, value standard deviations remain relatively small across channels. The channel-wise standard deviation indicates the range of numerical distribution across channels, corresponding to $\boldsymbol{X}{max} - \boldsymbol{X}{min}$ in Fig.~\ref{eq:quant}. This suggests that integer quantization of keys is more challenging than values, necessitating distinct quantization precisions for different channels.

Given the presence of amplitude and standard deviation outliers in KV cache, we propose using PQ for KV cache compression. PQ splits vectors into multiple subspaces and clusters multiple channels within each subspace for quantization. Channels in the same subspace are uniformly quantized to $2^{nbits}$ states, where $nbits$ denotes the quantization bits of each subspace vector. Following PQ centroid training principles, channels with higher quantization difficulty will occupy more states in the PQ centroid, inherently supporting mixed precision quantization between channels.

\section{Efficient KV Cache Compression via PQ}
\subsection{\archname Framework Overview}

\begin{figure}[t]
\setlength{\abovecaptionskip}{1pt}  
\setlength{\belowcaptionskip}{1pt} 
\centering
\includegraphics[width=0.95\linewidth]{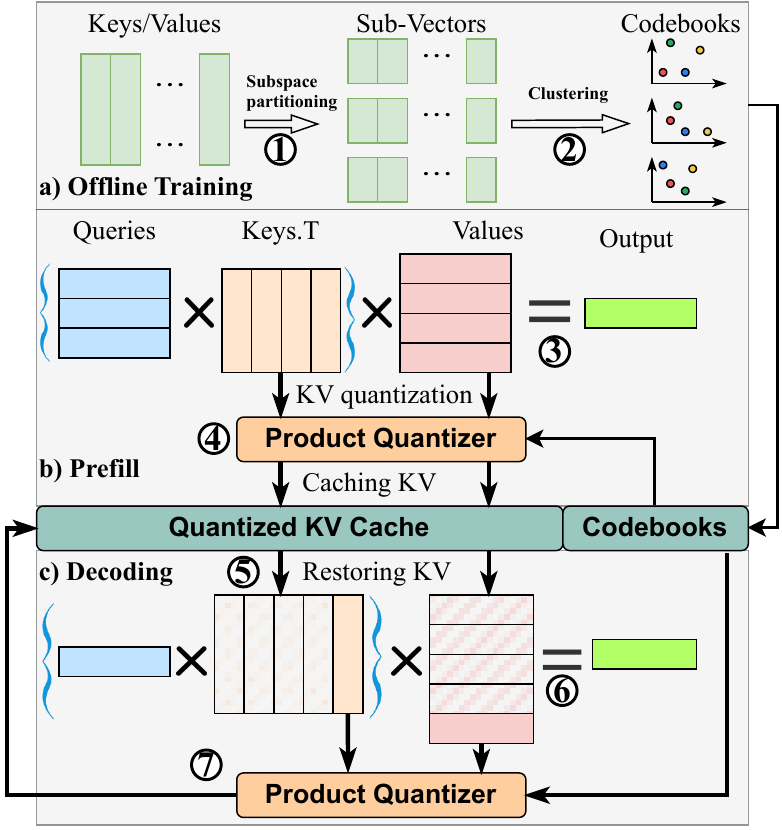}
\caption{An overview of \archname algorithm framework. \archname's algorithm framework consists of three parts: 1) offline PQ centroid training; 2) Prefill stage with KV quantization; 3) Decode stage with KV quantization.}
\label{fig:overview}
\end{figure}

This section illustrates the algorithm framework of the proposed \archname, which enhances dynamic KV compression using the PQ algorithm. As shown in \Cref{fig:overview}, \archname consists of three main phases: (1) offline PQ codebook training, (2) online prefill with KV cache quantization, and (3) online decode with KV cache quantization.

\textbf{Offline PQ Codebook Training.}
The process begins with offline PQ codebook training of the KV cache, similar to obtaining activation distribution in activation quantization scenarios. During baseline inference, KV cache is sampled, and then sampled keys and values are used for PQ training to determine the codebooks. \Cref{fig:overview} a) illustrates this process: sampled key/value vectors are first partitioned into multiple sub-vectors \hbquan{1}, then the k-means algorithm is applied to find centroids for each sub-space \hbquan{2}. After that, the trained centroids are stored in GPU memory as codebooks for future quantization.

\textbf{Online Prefill with KV Cache Quantization.}
During the prefill phase, the generated full-precision keys and values are used for attention computation \hbquan{3}, and quantized with the trained codebooks \hbquan{4}. The compressed KV (centroid indices) are then saved into the KV cache. The quantization and de-quantization process can be expressed as \cref{eq:kvquant} and \cref{eq:kvdequant}, respectively.

\begin{equation}
\label{eq:kvquant}
\left\{
\begin{array}{l}
    \Delta_K = \concat_{n=1}^{M} argmin(\lVert (K^i \times C_K^i) \rVert) \\
    \Delta_V = \concat_{n=1}^{M} argmin(\lVert (V^i \times C_V^i) \rVert)
\end{array}
\right.
\end{equation}

\begin{equation}
\label{eq:kvdequant}
\left\{
\begin{array}{l}
    \hat{K} = \concat_{n=1}^{M} \Delta_K^i \otimes C_K^i \\
    \hat{V} = \concat_{n=1}^{M} \Delta_V^i \otimes C_V^i
\end{array}
\right.
\end{equation}
where $M$ indicates the subspace number of PQ, $\concat$ denotes concatenation of M sub-vectors, $C_K^i$ and $C_V^i \in \mathbb{R}^{2^{nbits} \times d_k}$ are the $i$-th PQ codebooks of $K$ and $V$. $\Delta_K^i$ and $\Delta_V^i \in \mathbb{R}^{n_k}$ are the quantized $K$ and $V$ of the $i$-th subspace, and $nbits$ denotes the the quantization bits for each subspace of PQ, so the $2^{nbits}$ is the codebook size of each subspace, and the symbol $\otimes$ means column-wise sampling $C^i$ according to the indices of $\Delta_i$. $\hat{K}$ and $\hat{V}$ represent the de-quantized $K$ and $V$.

\textbf{Online Decode with KV Cache Quantization.}
In the decode phase, the current token generates pairwise query, key, and value vectors. Based on the auto-regressive mechanism, the current token needs to compute attention not only with itself but also with all historical tokens \hbquan{6}. This phase requires traversing all codes in the KV cache and restoring them to float vectors by looking up the codebooks. The attention computation can be represented as:
\begin{equation}
\label{eq:pqattn}
    A_n = \mathrm{Softmax}(\frac{q_n \times [\hat{K}_{n-1}^T, k_n^T]}{\sqrt{d_k}}) \times [\hat{V}_{n-1}, v_n]
\end{equation}
Where $A_n$ denotes the attention output of the $n$-th token.
We obtain $\hat{K}$ and $\hat{V}$ by \cref{eq:kvdequant}. Lastly, the current key and value are quantized using codebooks and appended to the KV cache\hbquan{7}.

\subsection{Efficient Quantization Transformation}
While PQ compression reduces KV cache memory demands, quantization and de-quantization overheads (as shown in \cref{eq:kvquant,eq:kvdequant,eq:pqattn}) still persist during the decoding phase. This section introduces an efficient design to address these inefficiencies.

\textbf{Attention Computation Transformation}.
We begin by transforming the attention computation as follows:
\begin{equation}
\label{eq:olattn}
\begin{aligned}
A_n = &\mathrm{OlSoftmax}(\frac{q_n \times \concat_{i=1}^M{C_K^i}^T \otimes {\Delta_K^i}^T}{\sqrt{d_k}}) \times \concat_{i=1}^M \Delta_V^i \otimes C_V^i \\
    &+ \mathrm{OlSoftmax}(\frac{q_n \times k_n}{\sqrt{d_k}}) \times v_n
\end{aligned}
\end{equation}

Here, \textit{OlSoftmax($\cdot$)} denotes an online softmax operation that splits the attention computation into two parts:
\begin{enumerate}
    \item The current token's attention output to past tokens (first term $\mathrm{OlSoftmax}(\frac{q_n \times \concat_{i=1}^M{C_K^i}^T \otimes {\Delta_K^i}^T}{\sqrt{d_k}}) \times \concat_{i=1}^M \Delta_V^i \otimes C_V^i$)
    \item The current token's self-attention output (second term $\mathrm{OlSoftmax}(\frac{q_n \times k_n}{\sqrt{d_k}}) \times v_n$)
\end{enumerate}

\makeatletter
\newcommand\figcaption{\def\@captype{figure}\caption}
\newcommand\tabcaption{\def\@captype{table}\caption}
\makeatother
\begin{figure*}[t]
    \vspace{-1.3cm}
    \setlength{\abovecaptionskip}{1pt}  
\setlength{\belowcaptionskip}{1pt} 
    \begin{minipage}[t]{0.4\textwidth}
    \centering
    \tabcaption{Detail model information for evaluation.}
    \label{tab:models}
    \resizebox{0.95\linewidth}{!}{
    \renewcommand{\arraystretch}{1.31}
        \begin{tabular}{@{}cccc@{}}
        \toprule
         & Parameters & \begin{tabular}[c]{@{}c@{}}positional\\[-3.5pt] embedding\end{tabular} & \begin{tabular}[c]{@{}c@{}}sequence\\ [-3.5pt] length\end{tabular} \\ \midrule
        GPT2-xl      & 1.5 B & Absolute & 1024 \\
        LLaMA-2-7B   & 7B    & RoPE     & 4096 \\
        MPT-7B       & 7B    & ALiBi    & 2048 \\
        Longchat-7B  & 7B    & RoPE     & 32K \\
        Yarn-LlaMA-2-7B & 7B    & RoPE     & 128K \\ \bottomrule
        \end{tabular}
    }
    \end{minipage}
    \hspace{-0.23cm}
    \begin{minipage}[t]{0.59\textwidth}
    \setlength{\abovecaptionskip}{1pt}  
    \setlength{\belowcaptionskip}{1pt} 
    \centering
    \tabcaption{Perplexity evaluation results on various models across different datasets.}
    \resizebox{0.93\linewidth}{!}{
    \begin{tabular}{@{}llllllll@{}}
        \toprule
        \multicolumn{1}{l|}{\multirow{2}{*}{Method}} & \multicolumn{2}{l}{GPT2-xl} & \multicolumn{3}{l}{LLaMA-2-7B} & \multicolumn{2}{l}{MPT-7B} \\ \cmidrule(l){2-8} 
        \multicolumn{1}{l|}{}                        & Wikitext-2       & PTB      & Wikitext-2    & PTB    & Wikitext-2      & PTB      \\ \midrule
        baseline       & 17.41  & 21    & 5.12  & 28.31    & 7.68     & 9.98  \\
        KVQuant-3b     & 18.59  & 23    & 11.21 & 12323.75 & 27681432 & 3E+07 \\
        KVQuant-3b-1\% & 18.24  & 22.35 & 5.22  & 24.34    & 7.826    & 10.14 \\
        MILLION-3b     & 17.6   & 21.14 & 5.2   & 29.55    & 7.79     & 10.08 \\
        KVQuant-4b     & 18.156 & 22.32 & 6.99  & 102.21   & 24043348 & 3E+07 \\
        KVQuant-4b-1\% & 18.108 & 22.15 & 5.14  & 25.86    & 7.71     & 10.02 \\
        MILLION-4b     & 17.57  & 21.12 & 5.21  & 29.56    & 7.75     & 10.03 \\ \bottomrule
    \end{tabular}
    }
    \label{tab:ppl}
    \end{minipage}%
    \vspace{-0.3cm}
\end{figure*}

The second part of the Eq. (\ref{eq:olattn}), dealing with the current token's self-attention, thus uses the full-precision key and value. It has $\mathcal{O}(1)$ computational complexity and requires no quantization or de-quantization for subsequent token generation.
The first part, on the other hand, addressing attention to past tokens, is computed through two steps:
1) Calculate $q_n \times \concat_{i=1}^M{C_K^i}^T$ with $\mathcal{O}(1)$ complexity; 2) Sampling the resulting vector with $\Delta_K$, having $\mathcal{O}(n)$ complexity.

\textbf{Efficient Attention Implementation.}
Given that $\Delta_K$ matrix columns are index vectors, we developed an efficient CUDA kernel. This kernel loads $q_n \times \concat_{i=1}^M{C_K^i}^T$ as a look-up table into GPU L1 cache for fast random accessing during sampling and obtains the final result through sampling, thus the quantized $\Delta_K$ can be computed directly without de-quantization. Besides, the $\Delta_K$ matrix is prefetched with float4 data format, which further enhances the efficiency of continuous memory access. For instance, with 8-bit subspace product quantization ($nbits$ = 8), each thread in the CUDA kernel can simultaneously load 16 indexes into registers in float4 format, and then use these to perform 16 shared memory address lookups in int8 format. 
Furthermore, the kernel effectively paralleliz table lookup of $\Delta_K$ and de-quantization of $\Delta_v$ over CUDA thread blocks inspired by flash decoding.

\textbf{Asynchronous Quantization Strategy.}
As of this point of introducing, quantization of $k_n$ and $v_n$ is still performed sequentially before the attention mechanism. However, cached codes are not needed until next token generation, indicating the computational independence of these two steps. Given the fact that decoding phase is memory-bounded, we assign quantization to a lower-priority CUDA stream to enable asynchronous execution of quantization and the subsequent processing, aiming at maximizing GPU utilization.

\subsection{\archname Dataflow}

\begin{figure}[t]
\setlength{\abovecaptionskip}{1pt}  
\setlength{\belowcaptionskip}{1pt} 
\centering
\includegraphics[width=0.92\linewidth]{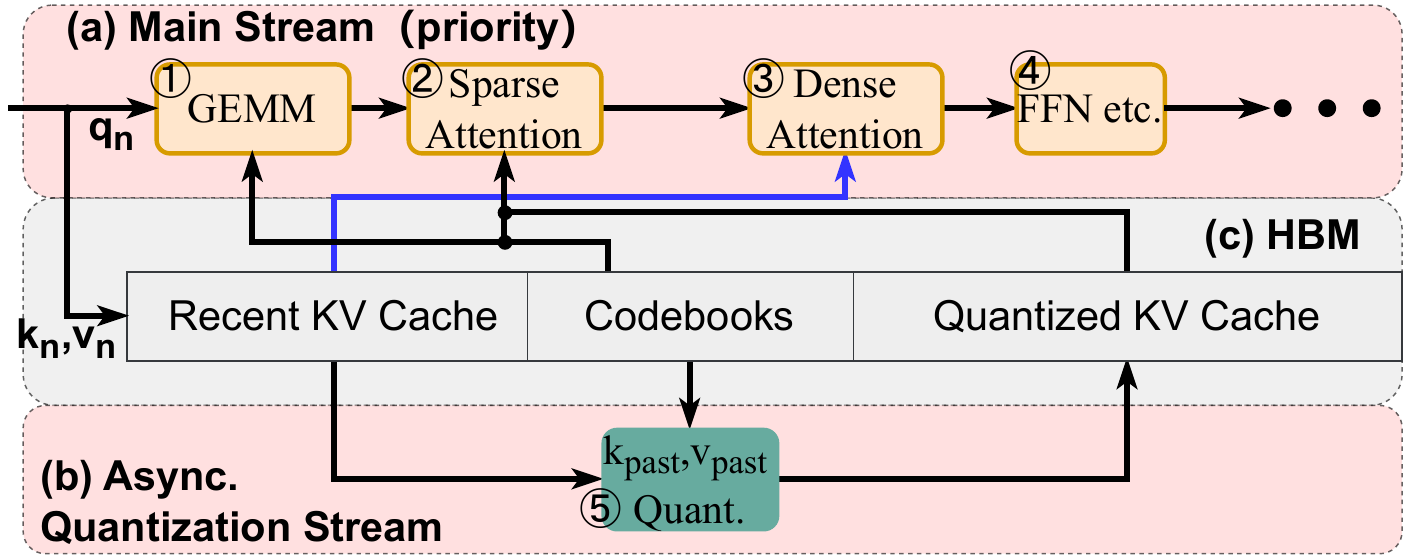}
\caption{Efficient Attention Implementation of \archname with Asynchronous Quantization.}
\label{fig:dataflow}
\vspace{-0.3cm}
\end{figure}

The dataflow of \archname for LLM decoding is illustrated in \cref{fig:dataflow}. To minimize the impact of (de-)quantization on inference performance, \archname employs two CUDA streams:
1) A high-priority main (CUDA) stream for attention computation;
2) A low-priority auxiliary (CUDA) stream is responsible for key and value quantization.

The process begins with the hidden state undergoing linear transformation to generate the query, key, and value for the current token, which serve as inputs to the attention layer. As analyzed in the previous section, the query performs a GEMM operation with the codebooks to calculate distances from each query to all centroids in the codebooks \hbquan{1}. The resulting distance vector is then used for sparse attention calculation with the quantized keys and values \hbquan{2}. Concurrently, the query undergoes dense attention computation with the recent KV cache (full precision) \hbquan{3}. Sparse and dense attention outputs are combined through online softmax and fed into subsequent operators for further inference \hbquan{4}. To avoid blocking the the main stream, keys, and values generated during decoding are temporarily stored in the recent KV cache. When GPU utilization of the main stream's CUDA kernel is low, the asynchronous quantization stream initiates the quantization kernel. It retrieves full-precision KV from the static recent KV cache for batch quantization \hbquan{5}. This design efficiently balances the computational demands of attention processing and quantization, optimizing overall inference performance.

\section{Experiments}


\subsection{Experimental Setup}

\textbf{Models.}
Given the critical role of positional embeddings in the query and key components of LLMs, we conduct extensive experiments across various models employing different positional encoding techniques. 
Detailed specifications of these models are presented in \cref{tab:models}.


\textbf{Tasks.}
We evaluate the quantization performance of MILLION across two distinct task categories: First, to assess model accuracy, we employ Perplexity (PPL), widely regarded as the most stringent metric, to compare the accuracy of various quantization schemes during the prefill stage. We conduct the PPL experiments on both Wikitext-2 and PTB datasets. Second, we conduct generation tasks using LongBench \cite{bai2023longbench} for long context scenarios. This comprehensive assessment covered a range of linguistic challenges across various context lengths.

\begin{figure*}[t]
\vspace{-1.3cm}
\setlength{\abovecaptionskip}{1pt}  
\setlength{\belowcaptionskip}{1pt} 
\centering
\includegraphics[width=0.85\linewidth]{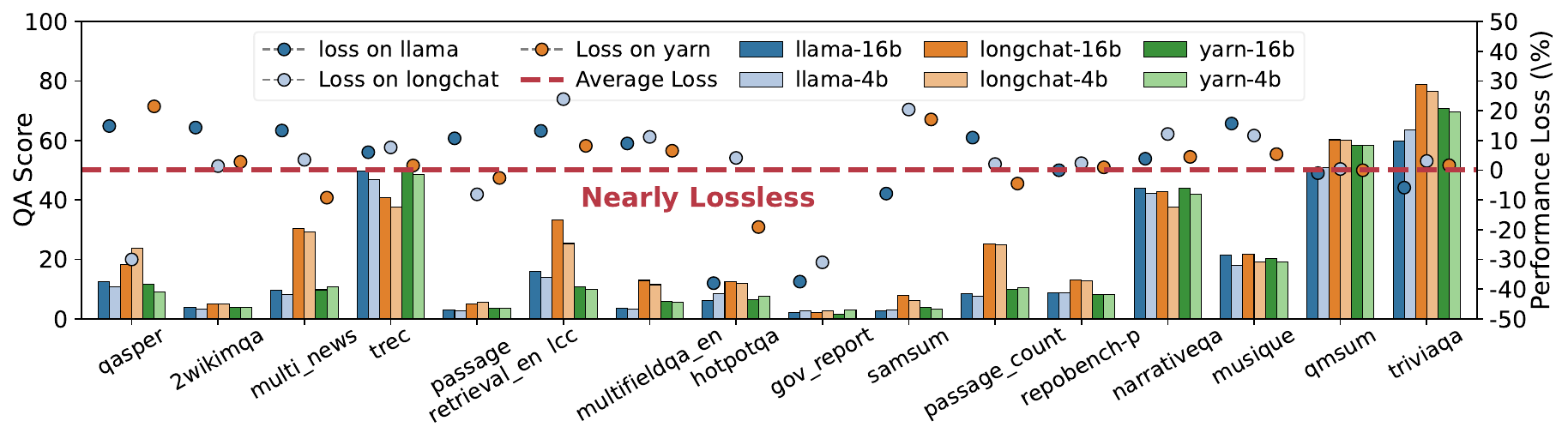}
\vspace{-0.4cm}
\caption{Performance evaluation of \archname on various models across a range of benchmarks in LongBench. While 16b denotes the baseline performance with fp16 KV cache, and the 4b depicts the \archname performance of 4 bits KV quantization.
}
\label{fig:qa}
\vspace{-0.2cm}
\end{figure*}
\textbf{Baselines.}
We use the LLM model in float16 format as the baseline. We also compare our method with two state-of-the-art KV cache quantization works: KIVI\cite{liu2024kivi} and KVQuant\cite{hooper2024kvquant10millioncontext}. For these comparisons, we use the quantization bit widths recommended in their original papers. KIVI employs 2-bit and 4-bit quantization, while KVQuant adopts 3-bit and 4-bit quantization.\footnote{We conclude that the PPL of KIVI is huge and therefore omit it from the PPL experiment.}


\subsection{Accuracy Results}


\textbf{PPL Evaluation.}
\cref{tab:ppl} compares the perplexity between KVQuant and \archname. We reproduce KVQuant's evaluation on the Llama-2 model and extended the comparison to various models and datasets. KVQuant uses non-uniform quantization for all configurations in \cref{tab:ppl}.
The results show that KVQuant, despite combining key/value quantization in different dimensions with non-uniform quantization, still incurs significant PPL loss. In contrast, full-precision sparse storage of 1\% outliers achieves near-lossless performance at 3 bits. This demonstrates the impact of outliers on low-bit-width quantization.
While isolating outliers for mixed-precision quantization effectively reduces KV cache quantization bit-width, it has drawbacks. This approach leads to unaligned memory access and introduces performance bottlenecks due to online sparse encoding, decoding, and computation.
\archname, employing uniform quantization, outperforms KVQuant in both 4-bit and 3-bit quantization\footnote{We scanned different combinations of $M$ and $nbits$ of PQ, and in this paper, the experiments adopted the best accuracy conditions of 3b and 4b, which were (64, 8) and (32, 12), respectively.}. Its accuracy is slightly lower than KVQuant's 1\% outlier mixed-precision scheme. This indicates that \archname effectively handles low-bit-width quantization with outliers while maintaining scheme efficiency.

\textbf{LongBench Evaluation.}
\Cref{fig:qa} shows \archname performance on various models across LongBench benchmarks, the residual block size is set to 0 for stress evaluation. While 4-bit quantization shows some performance degradation compared to full-precision 16-bit models, it maintains sufficient accuracy for many applications. Specifically, \archname with 4-bit KV cache quantization resulted in a score drop of 0.95 for llama-2-7b and 0.93 for longchat-7b. Notably, \archname improved the score of yarn-llama-2-7b (with a $128K$ context) by 0.45, demonstrating its effectiveness for models with the longest contexts.

\textbf{Outlier-immune Study.}
To verify MILLION's sensitivity\footnote{Sensitivity represents the proportion of PPL reduction caused by considering 1\% sparse outliers.} to outliers, we conduct additional experiments, as shown in Table~\ref{tab:outlierppl}. We identified the top 1\% of outliers in the KV cache based on data distribution and stored them in a sparse full-precision format. After applying MILLION to the filtered KV cache on LLMs, the PPL improved by only -0.38\% and 0.58\% in the 3-bit and 4-bit quantization scenarios, respectively. This marginal improvement suggests that MILLION is naturally robust to outliers, and incorporating sparse outlier handling offers little benefit, \emph{while significantly degrading end-to-end system performance.}

\begin{table}[]
\setlength{\abovecaptionskip}{1pt}  
\setlength{\belowcaptionskip}{1pt} 
\centering
\caption{The influence of 1\% outliers on PPL on KVQuant and MILLION}
\label{tab:outlierppl}
\begin{tabular}{@{}lllllll@{}}
\toprule
 &
  kv-3b &
  \begin{tabular}[c]{@{}l@{}}kv-3b\\ -1\%\end{tabular} &
  \begin{tabular}[c]{@{}l@{}}sensitivity\\ -3b\end{tabular} &
  kv-4b &
  \begin{tabular}[c]{@{}l@{}}kv-4b\\ -1\%\end{tabular} &
  \begin{tabular}[c]{@{}l@{}}sensitivity\\ -4b\end{tabular} \\ \midrule
KVQuant &
  11.21 &
  5.22 &
  {\color[HTML]{FE0000} \textbf{53.4\%}} &
  6.99 &
  5.14 &
  {\color[HTML]{FE0000} \textbf{26.5\%}} \\
MILLION &
  5.2 &
  5.22 &
  {\color[HTML]{3531FF} \textbf{-0.38\%}} &
  5.21 &
  5.18 &
  {\color[HTML]{3531FF} \textbf{0.58\%}} \\ \bottomrule
\end{tabular}
\vspace{-0.3cm}
\end{table}

\subsection{System Performance}

\begin{table}
\setlength{\abovecaptionskip}{1pt}  
\setlength{\belowcaptionskip}{1pt} 
\centering
\caption{TPOT (ms) comparison of KV quantization methods.}
\label{tab:tpotcomp}
\begin{tabular}{@{}lllllll@{}}
\toprule
Prefill Length  & 1K    & 2K    & 4K    & 8K    & 16K   & 32K    \\ \midrule
Baseline(fp16)  & 32.53 & 35.64 & 42.04 & 54.83 & 80.49 & 132.97 \\
KIVI(4b)        & 46.69 & 46.88 & 46.92 & 47.86 & \textcolor{red}{OOM}     & \textcolor{red}{OOM}      \\
KVQuant(4b)     & 75.73 & 73.92 & 75.34 & 74.90 & 78.17 & 90.16   \\
MILLION(4b)     & \textcolor{blue}{30.36} & \textcolor{blue}{31.57} & \textcolor{blue}{34.05} & \textcolor{blue}{38.34} & \textcolor{blue}{46.53} & \textcolor{blue}{63.41}  \\ \bottomrule
\end{tabular}
\vspace{-0.3cm}
\end{table}

\textbf{Generation Speed:}
We compare the performance of different KV quantization schemes by running Llama2-7b inference tasks on an A40 GPU. 
We assessment the average time per output token (TPOT) across different prefill lengths with 100 tokens generation.
All schemes use 4-bit KV quantization, with results shown in \cref{tab:tpotcomp}. KIVI, which uses a simpler quantization scheme, outperforms the baseline at 8K context lengths. In contrast, KVQuant’s non-uniform quantization significantly slows inference, and the outlier sparsity scheme further reduces performance. MILLION reduces computational complexity while maintaining high compression, delivering superior performance even at 1K context length.

\begin{figure}[t]
\setlength{\abovecaptionskip}{1pt}  
\setlength{\belowcaptionskip}{1pt} 
\centering
\includegraphics[width=0.95\linewidth]{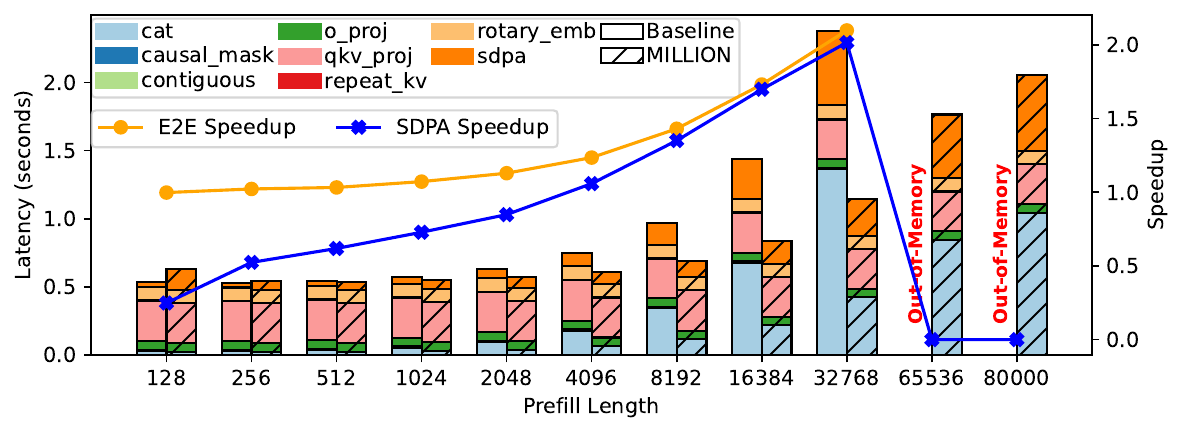}
\vspace{-0.3cm}
\caption{Latency breakdown analysis. MILLION significantly outperforms the baseline on two KV access operators, the performance margins increasing with the severity of the memory bottleneck.}
\vspace{-0.3cm}
\label{fig:lat_brk}
\end{figure}

\textbf{Latency Breakdown:}
\cref{fig:lat_brk} shows the latency breakdown of each operator to highlight the performance gains in the MILLION system. Compared to the baseline, MILLION achieves significant performance improvements in two operators: Scaled Dot Product Attention (SDPA), which performs the attention operation, and Concatenate (cat), which manages the KV cache. This demonstrates that MILLION suppresses the access bottleneck in the generation phase by compressing the KV bitwidth, while reducing the computational complexity of attention further enhances performance. In contrast, other compression schemes incur higher computational overhead due to de-quantization. Besides, \cref{fig:lat_brk} shows the SDPA and end-to-end performance gains at different context lengths. The speedup of MILLION increases with context length, reaching 2.01x and 2.09x in attention and end-to-end acceleration ratios, respectively, at 32K context length. 

\section{Conclusion}
We present MILLION, a novel and efficient PQ-based quantization framework for LLM inference. \archname efficiently compresses data while preserving accuracy. It incorporates asynchronous quantization, enhancing inference speed and reducing additional operations caused by hardware-inefficient quantization/dequantization for KV cache compression. Experiments demonstrate that \archname achieves better model performance while maintaining low latency.




\bibliographystyle{IEEEtran}
\bibliography{references}

\end{document}